\documentclass[aps, twocolumn, superscriptaddress, reprint, nobalancelastpage]{revtex4-1}

\usepackage{amsmath}
\usepackage{amssymb}
\usepackage{mathtools}
\usepackage{braket}
\usepackage{txfonts}
\usepackage{graphicx}
\graphicspath{{figures/}}

\usepackage{xcolor}

\usepackage[colorlinks,urlcolor=blue,linkcolor=blue,citecolor=blue]{hyperref}

\usepackage{tikz}
\usetikzlibrary{shapes}
\definecolor{mygreen}{HTML}{3AAB46}
\definecolor{myred}{HTML}{E42221}
\definecolor{myblue}{HTML}{314C9B}
\newcommand{\markerBlue}{{\tikz{\node[scale=0.65,circle,fill=myblue](){};}}}
\newcommand{\markerRed}{{\tikz{\node[scale=0.65,circle,fill=myred](){};}}}
\newcommand{\markerGreen}{{\tikz{\node[scale=0.65,circle,fill=mygreen](){};}}}
\newcommand{\markerTriangle}{{\tikz{\node[scale=0.4,regular polygon, regular polygon sides=3,fill=myred,rotate=0](){};}}}
\newcommand{\markerSquare}{{\tikz{\node[scale=0.6,regular polygon, regular polygon sides=4,fill=myred](){};}}}

\newcommand{\eqdef}{\mathrel{\mathop:}=}

\begin{document}

\title{Twins Percolation for Qubit Losses in Topological  Color Codes}
\author{D. Vodola}
\affiliation{Department  of  Physics,  Swansea  University,  Singleton  Park,  Swansea  SA2  8PP,  United  Kingdom.}

\author{D. Amaro}
\affiliation{Department  of  Physics,  Swansea  University,  Singleton  Park,  Swansea  SA2  8PP,  United  Kingdom.}

\author{M.A. Martin-Delgado}
\affiliation{Departamento de F\'isica Te\'orica I, Universidad Complutense, 28040 Madrid, Spain.}

\author{M. M{\"u}ller}
\affiliation{Department  of  Physics,  Swansea  University,  Singleton  Park,  Swansea  SA2  8PP,  United  Kingdom.}

\begin{abstract}
We establish and explore a new connection between quantum information theory and classical statistical mechanics by studying the problem of qubit losses in 2D topological color codes. We introduce a protocol to cope with qubit losses, which is based on the identification and removal of a twin qubit from the code, and which guarantees the recovery of a valid three-colorable and trivalent reconstructed color code. Moreover, we show that determining the corresponding qubit loss error threshold is equivalent to a new generalized classical percolation problem. We numerically compute the associated qubit loss thresholds for two families of 2D color code and find that with $p = 0.461 \pm 0.005$ these are close to satisfying the fundamental limit of 50\% as imposed by the no-cloning theorem. Our findings reveal a new connection between topological color codes and percolation theory, show high robustness of color codes against qubit loss, and are directly relevant for implementations of topological quantum error correction in various physical platforms.
\end{abstract}
\maketitle

Quantum information theory, widely recognised as a powerful paradigm to formulate and address problems in information processing beyond the realms of classical physics, has shown strong cross-connections to different fields, including atomic, molecular and optical (AMO) physics~\cite{ladd-nature-464-45}, condensed matter~\cite{Lewenstein2007,Amico2008}, computer science \cite{nielsen-book}, but also classical statistical mechanics. Exploring the connection between quantum information and classical statistical physics has proven particularly fruitful in both directions and revealed deep and unexpected links. For instance, efficient quantum algorithms enable estimating partition functions of classical spin systems~\cite{Lidar1997,Somma2007,Nest2007,Geraci2008,Cuevas2009,Geraci2010,Cuevas2011,Xu2011,Cuevas2016,Zarei2017}. In the context of fault-tolerant quantum computing, topological quantum error correcting (QEC) codes, such as Kitaev's surface code \cite{Kitaev2003, Dennis2002} and color codes~\cite{Bombin2006, bombin-prl-98-160502}, protect quantum information in two- or higher-dimensional lattices of qubits. They provide to date, arguably, the most promising route towards practical fault-tolerant quantum computers \cite{Terhal-RevModPhys.87.307}. Here, the problem of studying the error robustness of these topological quantum codes can be mapped onto classical statistical mechanics lattice models \cite{Dennis2002}, opening a powerful avenue to study fundamental features of the corresponding QEC codes.

For instance, error thresholds and the parameter regimes where QEC succeeds/fails, are identified with the critical point and ordered/disordered phases of the classical models, respectively. Depending on the quantum code and error model considered, different classical models emerge: For computational errors only, such as uncorrelated bit and phase flips, QEC can be mapped to a classical 2D random-bond Ising model with two-body interactions for the toric code~\cite{Dennis2002} and three-body interactions for the color code~\cite{Katzgraber2009}. If measurements in the QEC procedure are also faulty, the QEC problem maps for the toric code onto a classical 3D random plaquette gauge model~\cite{Dennis2002, Ohno2004} and for color codes onto a 3D lattice gauge theory, introduced for the first time in Ref.~\cite{Andrist2011}.

\begin{figure}
\centering
\includegraphics[width=0.425\textwidth]{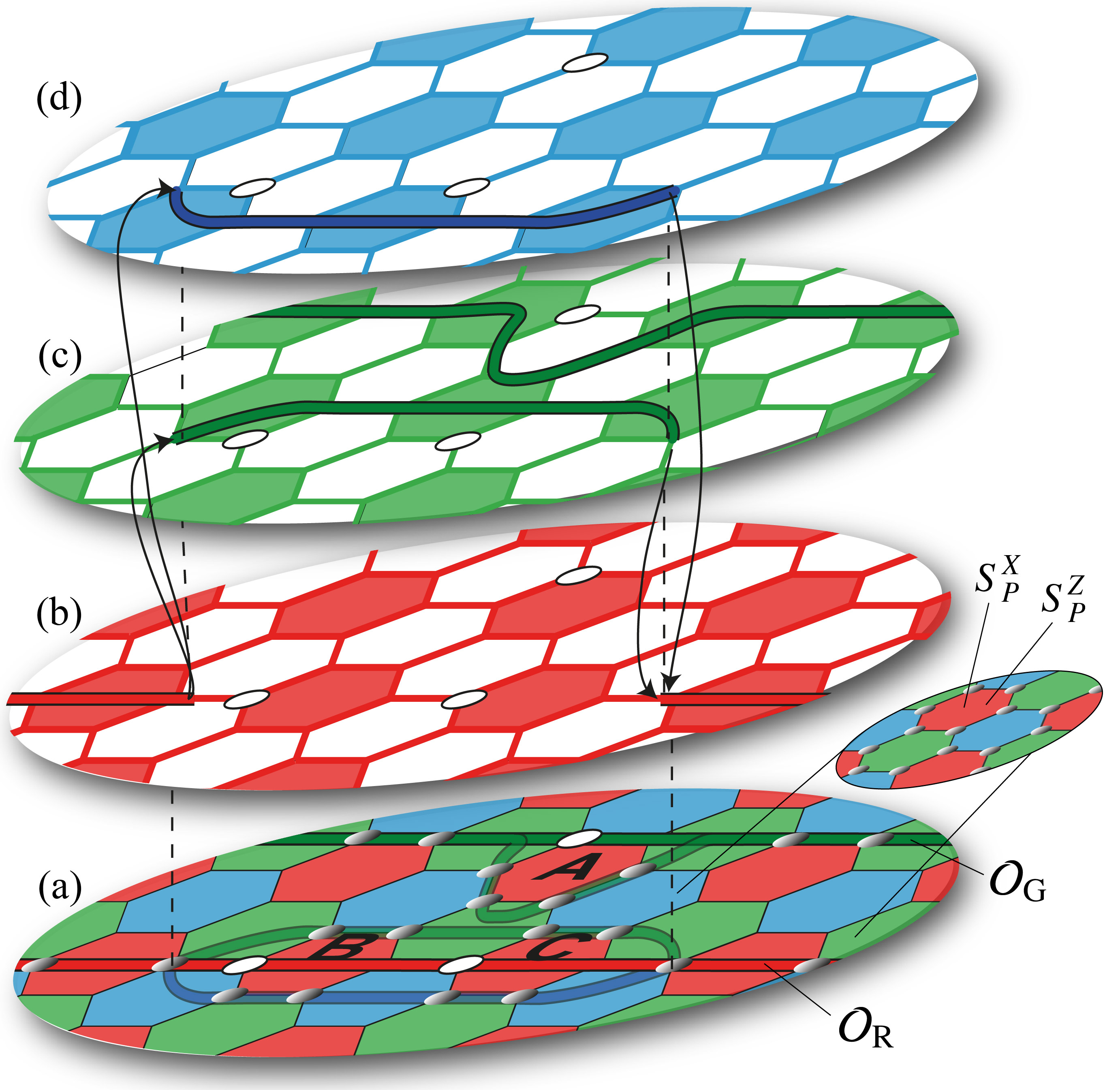}
\caption{(a) An excerpt of a hexagonal 2D color code. Physical qubits are located at the sites (vertices) of the lattice, and each plaquette~$P$ hosts an $X$ and $Z$-type generator $S^X_P$, $S^Z_P$ of the stabiliser group of the code. Logical (string) operators $\mathcal{O}_G$ and $\mathcal{O}_R$ can be deformed by the action of stabilizers, to evade locations where qubits are lost from the lattice (white circles): for example, $\mathcal{O}_G$ is deformed by the generator A into the lighter green path, whereas the red string $\mathcal{O}_R$ branches into an equivalent green and blue string by the action of generators B and C. For clarity, only qubits on which $\mathcal{O}_G$ and  $\mathcal{O}_R$ have support, are shown. Note that the branched red operator belongs to all three shrunk lattices of plaquettes as shown in panels (b,c,d).}
\label{fig:colorcode}
\end{figure}

Qubit loss, caused by actual loss of particles or photons, or by leakage processes that take the qubit out of the computational space, is an additional severe error source in many physical platforms, with some counterstrategies developed~\cite{grassl-pra-56-33,ralph-prl-95-100501,lu-pnas-105-11050,Sherman2013,Mehl2015}. For the surface code affected by qubit losses, correction of losses is related to a classical bond percolation transition on a square lattice~\cite{Stace2009, Stace2010}. For topological color codes, on the contrary, to date it is an open question (i) how to cope with qubit losses, (ii) if and to what classical model the problem of qubit loss correction can be mapped, and (iii) what level of robustness against losses color codes offer. 

In this Letter, we address the problem of qubit losses in topological color codes by (i) introducing an explicit novel protocol (algorithm) to correct losses, (ii) establishing a mapping of QEC color codes affected by losses onto a novel model of classical percolation and (iii) exploiting this mapping to compute the fundamental qubit loss error threshold in color codes associated with our protocol. Our results not only establish a new connection with classical percolation, but are also directly relevant for practical QEC and quantum computing, including with trapped ions~\cite{Chiaverini2004, Schindler2011, Lanyon2013, Nigg2014, Bohnet2016}, Rydberg atoms~\cite{isenhower-prl-104-010503, wilk-prl-104-010502, saffman-rmp-82-2313}, photons~\cite{OBrien2005, Pittman2005} and superconductors~\cite{Barends2014, Kelly2015, Corcoles2015, Ofek2016}.

\begin{figure*}\centering
\includegraphics[width=0.9\textwidth]{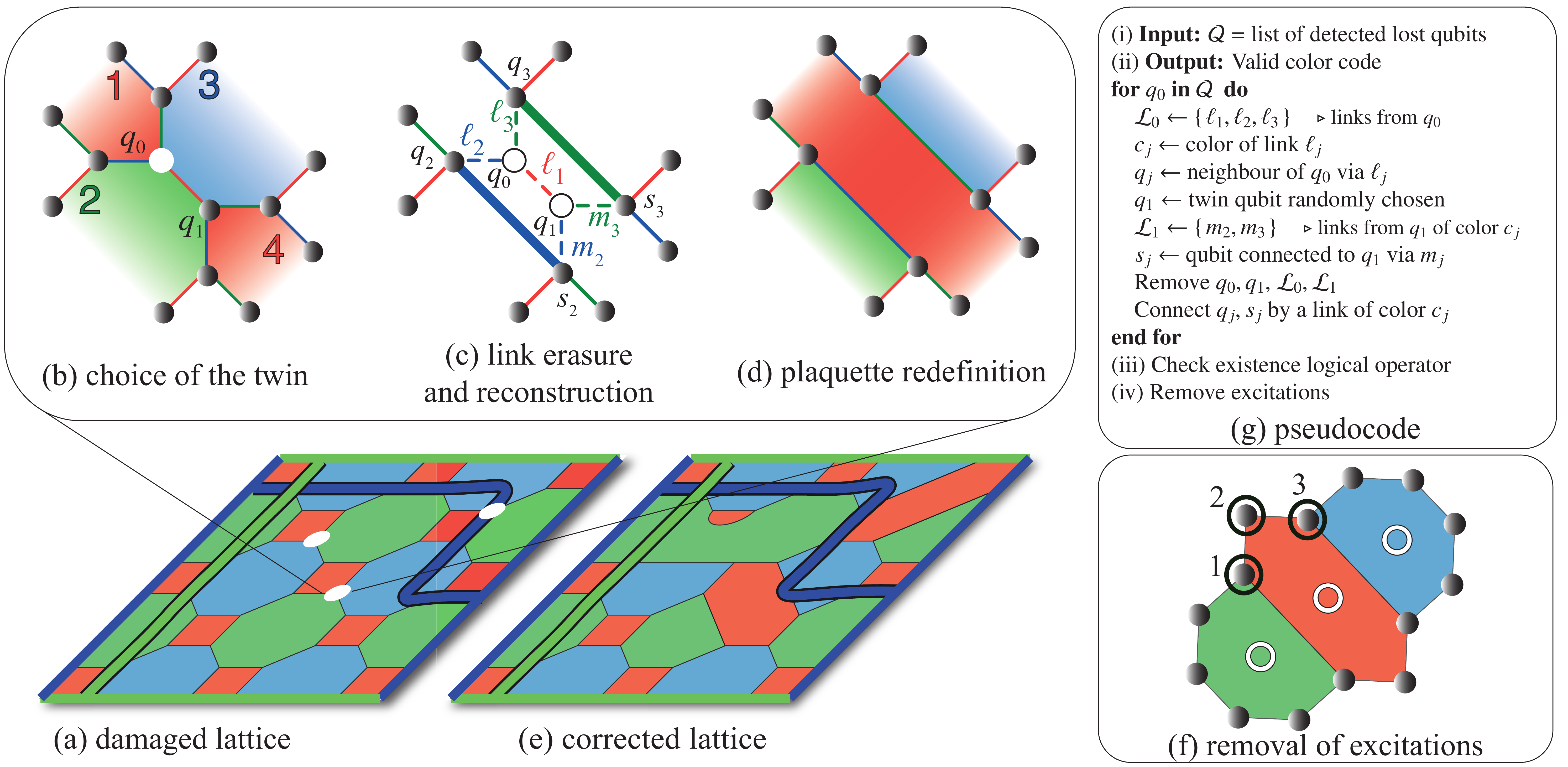}
\caption{(a) Representation of a distance $d = 6$ color code, on a 4.8.8 lattice, affected by three qubit losses (white circles). This code stores two logical qubits, with four logical operators defined along green and blue strings that span the lattice and connect opposite borders of the corresponding colors. Panels (b-d): Protocol to recover from losses in a color code (see also main text). (b) Twin identification. (c) Link removal and lattice reconstruction. (d) Plaquette redefinition. (e) Reconstructed lattice after loss correction. A logical operator (e.g. the blue one) that is potentially affected by a loss can be transformed into an equivalent one by multiplication with an operator from $\mathcal{S}$. (f) Removal of residual excitations (white hollow circles) of re-defined stabilizers by a Pauli corrections (open black circles) on a string of qubits (labelled 1,2,3). (g) Pseudocode summarizing the main steps (b,c) of the loss recovery protocol.}\label{fig:stabilizers_redefined}
\end{figure*}

\emph{The color code}~\cite{Bombin2006} is a topological QEC code defined by a stabilizer group $\mathcal{S}$ acting on physical qubits placed at the vertices of a trivalent, three-colorable 2D lattice [see Fig.~\ref{fig:colorcode}(a)]. Associated to each face or plaquette~$P$, two types of generators of $\mathcal{S}$ are defined as $S^{\sigma}_P =  \prod_{j\in  P} \sigma_j $, where $\sigma$ is a Pauli $X$ or $Z$ operator acting on all qubits $j$ belonging to $P$. For any pair of plaquettes $P,P'$, the corresponding generators $S^{X}_P$ and  $S^{Z}_{P'}$ mutually commute. 
As the lattice is three-colorable, one can associate one color $c$ (among $R,G,B$) to each plaquette and hence to each generator such that if two plaquettes share an edge they will be of different color. Each plaquette will also belong to one shrunk colored sub-lattice like in Fig.~\ref{fig:colorcode}(b,c,d)~\cite{sup_mat}. The code space $\mathcal{C}$, i.e.~the Hilbert sub-space hosting logical states, is the simultaneous +1 eigenstate of all $S^{\sigma}_P$ stabilizers. The number of logical qubits $k$ depends on the topology of the manifold in which the system is embedded, namely for a surface of Euler characteristic $\chi$ one finds $k = 4 - 2 \chi$~\cite{Bombin2006}. A special set of operators acting within $\mathcal{C}$ are the logical operators, $T^\sigma_\mu$, for $\mu = 1,\dots,k$ and $\sigma=X,Z$. %They commute with each of the stabilizer generators $S^{\sigma}_P$ but they do not belong to $\mathcal{S}$.
Their definition depends on the topology, e.g.~for a torus they are associated with the two nontrivial cycles around it~\cite{Lidar2013, sup_mat}.

\emph{The protocol for losses} -- A valid scheme that can correct for losses, in general, requires (i) detecting the lost qubits;
(ii) re-defining the set of stabilizer generators such that each of them has support only on qubits not affected by loss; (iii) checking if the encoded logical quantum states are unaffected by the losses; (iv) finally removing possible excitations ($-1$ eigenstates of newly defined stabilizer generators). Assuming the positions of the losses have been determined (i), we will now focus on the key steps (ii)-(iii) of the protocol and refer for details on (iv) to the Supplemental Material~\cite{sup_mat}.

\emph{Redefinition of stabilizers} -- For color codes one main challenge is (ii) to re-define a modified set of stabilizers respecting the constraint that the resulting modified lattice hosting the color code remains three-colorable and trivalent. The protocol we propose to achieve this is as follows: (1) given a detected lost qubit $q_0$ [depicted as a white circle in Figs.~\ref{fig:stabilizers_redefined}(a,b)] we choose randomly a \textit{twin qubit} $q_1$ among its three neighboring qubits. This twin qubit will be \textit{sacrificed}, i.e.~also removed from the code [Fig.~\ref{fig:stabilizers_redefined}(c)]. We refer to the pair of qubits $q_0$ and $q_1$ and the link connecting them as a \textit{dimer}. Note that the dimer connects two plaquettes of the same color [plaquettes 1 and 4 in Fig.~\ref{fig:stabilizers_redefined}(b)], it is shared by two neighbouring plaquettes of the two complementary colors [plaquettes 2 and 3 in Fig.~\ref{fig:stabilizers_redefined}(b)] and it is connected to four links $\ell_2, \ell_3, m_2,m_3$ (two pairs for each of the complementary two colors), see dashed lines in Fig.~\ref{fig:stabilizers_redefined}(c). (2) The dimer, as well as the two pairs of links originating from it are removed. Then, to redefine a valid trivalent and three-colorable lattice, the two pairs of qubits with links of the same colors as the removed ones are connected by new links [Fig.~\ref{fig:stabilizers_redefined}(c)]. In this way the two plaquettes originally connected by the dimer will merge into a single larger plaquette, while the two plaquettes that were sharing the dimer will shrink and have a qubit number reduced by two [Fig.~\ref{fig:stabilizers_redefined}(d)]. Steps (1) and (2) are repeated iteratively for all lost qubits. The final lattice is guaranteed to be trivalent and three-colorable, see Fig.~\ref{fig:stabilizers_redefined}(e) for an example. The quantum code reconstruction algorithm is summarised in Fig.~\ref{fig:stabilizers_redefined}(g). The validity of our protocol can be substantiated by computing the Euler characteristic of the resulting lattice. Before the occurrence of a loss, $\chi = V - E + F$ where $V,E$ and $F$ denote the numbers of qubits, links and plaquettes of the lattice. After (1) and (2), the Euler characteristics remains unchanged, as $\chi' = (V - 2 ) - (E - 3) + (F - 1) = \chi$, and so does consequently the number of encoded logical qubits.

 % For the purpose of discussing the recovery from qubit losses it is convenient to view the logical operators as product of Pauli operators [see, e.g., the green operator in Fig.~\ref{fig:colorcode}(a)] acting on strings spanning the entire lattice and made up by pairs of physical qubits on links that connect plaquettes of the same color. These string operators of different color thus belong to the three shrunk lattices of plaquettes . 
 
\emph{Check of the existence of the logical operators} -- In order to understand whether or not a given set of losses affects the logical quantum information, one has to verify whether the logical operators remain intact. The support of logical operators is not unique~\cite{sup_mat}; thus one can obtain equivalent logical operators $\tilde{T}_\mu^{\sigma}$ by multiplying an original one, $T_\mu^{\sigma}$, by any element of $\mathcal{S}$. This equivalence can be used to check if a logical operator is still defined by considering that it is possible to recover the $\mu$-th logical qubit, if one can find a subset $\mathcal{V} \subseteq \mathcal{S}$ such that the modified logical operator 
\begin{equation}\label{eqn:ModifiedLogicalOp}
\tilde{T}^\sigma_\mu = T^\sigma_\mu \prod_{S^\sigma_P \in \mathcal{V}}S^\sigma_P
\end{equation}
does neither have support on losses nor on twin qubits. If that is not possible, $T^\sigma_\mu$ is in an undefined state and the encoded quantum information corrupted.

We consider three ways of checking if logical operators remain unaffected by the losses and code reconstruction: 

(I) The first one is based on the fact that logical operators can take the form of non-trivial colored strings spanning the entire lattice, like, e.g., the green logical operator $\mathcal{O}_\text{G}$ in Fig.~\ref{fig:colorcode}(a). If $\mathcal{O}_\text{G}$ is multiplied by the stabilizer of the red plaquette A, it is deformed into the string operator with support on the lighter green path, but it will continue to belong to the green shrunk lattice [Fig.~\ref{fig:colorcode}(c)]. Thus, one can ask if it is possible to find \emph{percolating strings} in the shrunk lattices that do not have support on losses nor twin qubits. This is equivalent to finding a logical operator $\tilde{T}^\sigma_\mu$ as a solution of Eq.~\eqref{eqn:ModifiedLogicalOp} under the constraint of choosing elements of the subset $\mathcal{V}$ of the original $\mathcal{S}$ only among the stabilizers of the two colors that are complementary to the color of $T^\sigma_\mu$. Note that since one uses the original group $\mathcal{S}$ to find equivalent logical operators, these strings have support on chains made up by links, which do not involve losses nor twin qubits and which belong to the original lattice.

(II) The second way is by considering that the subset $\mathcal{V}$ can be formed also by stabilizers of the same color as $T^\sigma_\mu$. If, e.g., a red logical operator $\mathcal{O}_\text{R}$ [Fig.~\ref{fig:colorcode}(a)] is multiplied by stabilizers of two red plaquettes B and C, it will split into a green and blue string [Fig.~\ref{fig:colorcode}(b,c,d)]. This \emph{string branching} process, not present in the surface code~\cite{Kitaev2003, Dennis2002}, is peculiar to color codes, allowing a logical operator to take the form of a \textit{string-net}~\cite{Bombin2006}. The check for the existence is then translated into a combined percolation check taking place in the coupled three shrunk lattices. For, say, the operator $\mathcal{O}_\text{R}$, the starting point of such branching  [Fig.~\ref{fig:colorcode}(b)] is a qubit that has a red link where a loss or a twin qubit resides and the green and the blue unaffected links both belong to the original lattice. Then, the red operator can split up and percolate as a blue and a green string into the two shrunk lattices [Fig.~\ref{fig:colorcode}(c,d)]. The ending point of the branching is required to be a qubit having all the three unaffected links belonging to the original lattice. In this way, the blue and green strings can recombine into the red one that continues its way in the red shrunk lattice.

(III) The third way consists in solving Eq.~\eqref{eqn:ModifiedLogicalOp} directly. This approach allows for the most general equivalent logical operator, as it considers multiplication with elements from the whole original stabilizer group. Although the number of elements in $\mathcal{S}$ grows exponentially with the number of physical qubits, one can show that Eq.~\eqref{eqn:ModifiedLogicalOp} is a linear system of equations that can be efficiently solved~\cite{sup_mat}. Technique (III) defines the \emph{fundamental} limit $p_\text{fund}$ for the code as it captures the most general admissable forms of logical operators corresponding to percolating string-nets with in general several branching and fusing points at which the three coupled shrunk lattices supporting the string-net are coupled. 

The latter two methods (II) and (III) are a \textit{generalized percolation problem} that effectively deforms the operator by branching it into all three shrunk lattices. Before we discuss the resulting qubit loss error thresholds obtained by these three ways, it is instructive to briefly contrast our recovery protocol for color codes with the pioneering protocol for qubit losses in the surface code~\cite{Stace2009}. In the latter system, due to the absence of lattice constraints, two neighbouring plaquettes affected by the loss of a qubit located on the shared link can directly be fused into one larger plaquette (see~\cite{sup_mat}) that will host one new stabilizer with vanishing support on the lost qubit. Logical operators always remain string-type in nature and can be deformed so they evade the link corresponding to the lost qubit. Thus, the qubit loss problem in the surface code maps onto the bond-percolation problem on a 2D square lattice with associated threshold of 1/2~\cite{Stauffer1985}.

\emph{Removal of excitations} -- Returning to the color-code correction protocol, after steps (ii) and (iii), the new merged or shrunk $X$ and $Z$-type generators can show excitations ($-1$ eigenstates). If needed, these can be annihilated by physically applying a Pauli correction along a chain of qubits connecting the three new plaquettes, as shown in Fig.~\ref{fig:stabilizers_redefined}(f), or by a corresponding Pauli frame update on a software level~\cite{Aliferis2006,Knill2005}. This excitation removal, to re-initialize the QEC on the reconstructed lattice within the code space, does not require a decoder, is fully deterministic and iteratively applicable to more than a single qubit loss~\cite{sup_mat}.

\begin{figure}
\centering %
\includegraphics[width = 0.48\textwidth]{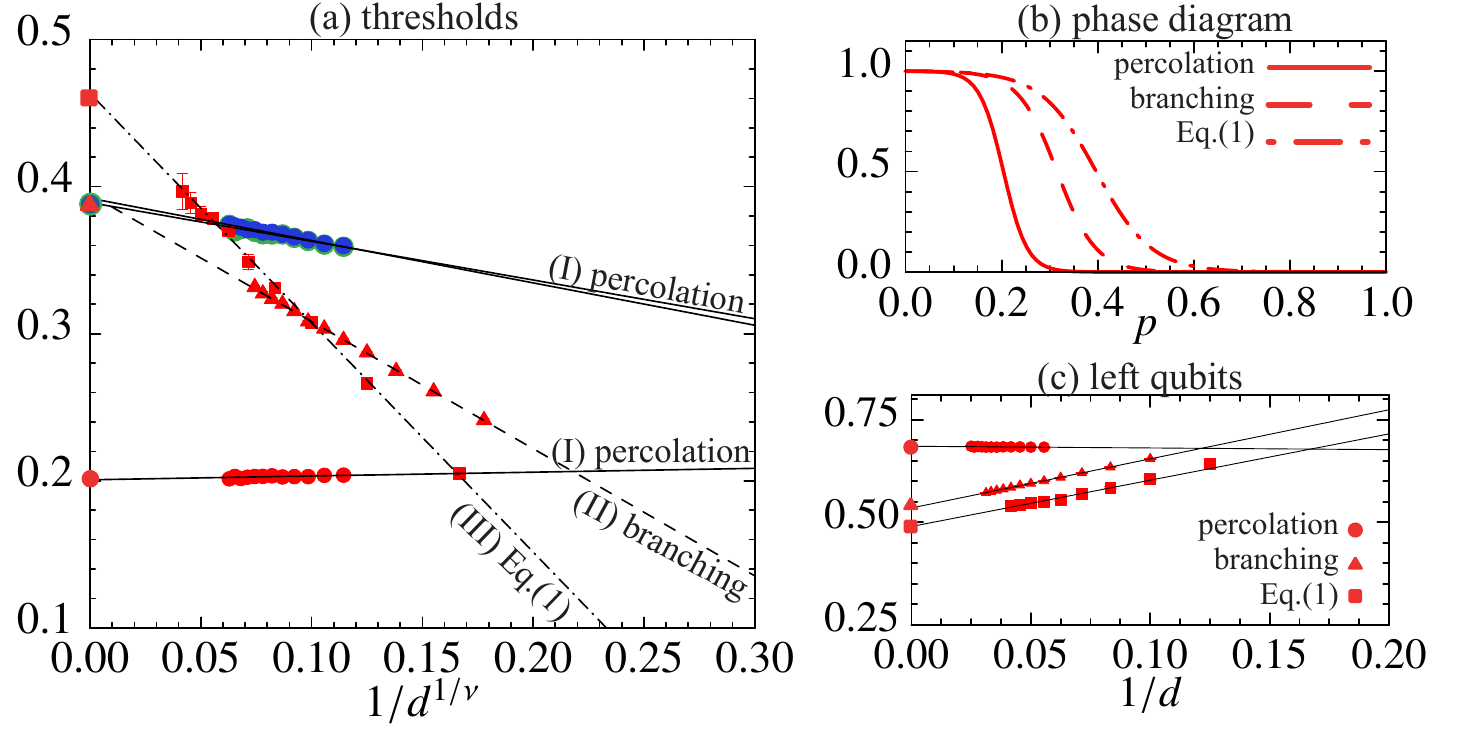}
\caption{(a) Loss thresholds for a 4.8.8 lattice computed by checking percolation for the three logical operators (\protect\markerBlue{} \protect\markerRed{} \protect\markerGreen), branching~(\protect\markerTriangle) and the existence of solutions of Eq.~\eqref{eqn:ModifiedLogicalOp} (\protect\markerSquare) only for the red logical operator. Thresholds are plotted as a function of $1/d^{1/\nu}$, with $d$ the distance of the code and $\nu = 4/3$. The intercepts of the black lines (marked with the same symbols as the data) represent the critical threshold in the thermodynamic limit. (b) Phase diagram at $d = 36$ showing the probability to find a logical operator using the three methods explained in the main text. (c) Fraction of qubits left in the lattice for each of the three methods as a function of $1/d$. For methods (II) and (III), the fractions approach the fundamental 50\% limit.}
\label{fig:NumericalResults}
\end{figure}

\emph{Numerical results} -- We now study the robustness of color codes against losses, using the above loss correction protocol. To this end, we consider 2D color codes of logical distances $d$ defined on planar 4.8.8 and a 6.6.6 lattices~\cite{sup_mat}. We generate random sequences of losses by Monte Carlo simulations, then reconstruct the lattice according to our protocol, and finally check (I) if percolating strings exist, (II) if percolating branched strings exist and (III) if the linear system of Eqs.~\eqref{eqn:ModifiedLogicalOp} admits solutions. In this way we determine the loss tolerance thresholds and map out the phase diagram which separates correctable and uncorrectable phases.

For the existence check of percolating strings (I) in a code of distance $d$, we compute, for each of the three colored shrunk lattices, the critical fraction of losses $p(d)$ at which, for the first time, a percolating string-type path ceases to exist~\cite{sup_mat}. Percolation theory~\cite{Stauffer1985} predicts the critical fraction $p(d)$ to scale in the limit $d\to\infty$ as $p(d)-p_\infty \propto 1/d^{1/\nu}$ with a scaling exponent~$\nu$. Numerically we find $\nu = 4/3$, which is the value expected from percolation theory~\cite{Stauffer1985}. Figure~\ref{fig:NumericalResults}(a) shows $p(d)$ and least-square linear fits whose intercepts for $1/d^{1/\nu} \rightarrow 0$ yield the string percolation thresholds $p_\text{perc}$ for each of the string-type logical operators of the three different colors. We obtain similar results for the 6.6.6 lattice~\cite{sup_mat}. Red string-type logical operators have a lower threshold as the structure of the supporting red shrunk lattice is different from the other two ones. However, if we allow for (II) string branching, i.e.,~the red logical operator can split up into a green and a blue path [as in Fig.~\ref{fig:colorcode}(a)], and we compute, for $d\to\infty$, the fraction $p_\text{branch}$ at which a logical operator ceases to exist on the reconstructed coupled three shrunk lattices, we obtain a value compatible with $p_\text{perc}$ for the green and blue operators [red triangles along the dashed line in Fig.~\ref{fig:NumericalResults}(a)]. This result indicates that at a loss rate for which a red string-type operator does no longer percolate on the red shrunk lattice, branching allows the logical operator to evade the non-percolating red shrunk lattice and continue to propagate and percolate on the green and blue lattices instead, thereby almost doubling its robustness, from $p_\text{perc} \approx 0.2$ to $0.4$. Finally, we apply method (III) that checks the existence of a solution to Eq.~\eqref{eqn:ModifiedLogicalOp}. As expected, this yields the highest threshold of $p_\text{fund} = (0.461 \pm 0.005)$ [Fig.~\ref{fig:NumericalResults}(a), dot dashed line, linear fit of the data with $\nu = 1$]. For a $d=36$ lattice (containing originally 2452 physical qubits), Fig.~\ref{fig:NumericalResults}(b) displays the probability of finding a red logical operator as obtained from the three methods (I-III). This represents a phase diagram showing the boundary separating regions of low and high qubit loss rates, where the logical qubit associated to a logical red operator can and cannot be recovered. 

A natural question is how many qubits are left in the lattice at the percolation threshold, beyond which the encoded logical information cannot be fully restored. For our protocol, for low loss rates $p \ll 1$, when losses are sparsely distributed over the lattice, the fraction of remaining qubits is given by $1-2p$ as for each loss one twin qubit is also removed. However, for larger $p$ the sets of losses and of twin qubits can have a non-empty intersection, as, e.g.,~one of the twin qubits could correspond also to a loss. Figure~\ref{fig:NumericalResults}(c) shows the fraction of qubits left in the lattice for each of the three methods (I-III) as a function of $1/d$. Notably, when considering methods (II) and (III), this number approaches 50\%, which is the fundamental limit as imposed by the no-cloning theorem for the capacity of a quantum erasure channel~\cite{Bennett1997}. This shows not only a high intrinsic robustness of color codes against qubit loss. Importantly, it also underlines the near-optimality of our recovery protocol, which requires identifying twin qubits and elimination of dimers to guarantee a valid color code on the reconstructed lattice, and which is furthermore based on a \textit{local} reconstruction protocol, not taking into account the \textit{global} configuration of losses.

\emph{Conclusions} -- In this Letter, we have introduced an operationally defined and efficient protocol to cope with qubit losses in color codes, which preserves the three-colorability of the resulting reconstructed 2D lattice. We have established a new mapping of QEC color codes affected by qubit loss onto a novel model of classical percolation on coupled lattices. Finally, by computing the resulting loss error thresholds, we have shown that color codes in combination with our qubit loss correction protocol are highly robust against losses, almost saturating the fundamental limit set by the no-cloning theorem. Beyond the scope of this work, the protocol discussed can be extended to also account for computational and measurement errors, to investigate the trade-off between qubit losses and other error sources. Furthermore, we hope that the cross-connection of the QEC problem with a new generalised percolation problem will stimulate further research that leverages tools and results from percolation theory to investigate the robustness of other topological QEC codes and many-body quantum phases of matter under loss of particles.

We acknowledge support by U.S. A.R.O. through Grant No. W911NF-14-1-010. The research is also based upon work supported by the Office of the Director of National Intelligence (ODNI), Intelligence Advanced Research Projects Activity (IARPA), via the U.S. Army Research Office Grant No. W911NF-16-1-0070. The views and conclusions contained herein are those of the authors and should not be interpreted as necessarily representing the official policies or endorsements, either expressed or implied, of the ODNI, IARPA, or the U.S. Government. The U.S. Government is authorized to reproduce and distribute reprints for Governmental purposes notwithstanding any copyright annotation thereon. Any opinions, findings, and conclusions or recommendations expressed in this material are those of the author(s) and do not necessarily reflect the view of the U.S. Army Research Office. M.A.M.D.~acknowledges support from Spanish MINECO grant FIS2015-67411, and the CAM research consortium QUITEMAD+, Grant No. S2013/ICE-2801. We acknowledge the resources and support of High Performance Computing Wales, where most of the simulations were performed, as well as the HPC-Universit\'e de Strasbourg. D.V. thanks M. Guti\'errez for fruitful discussions.

\clearpage
\pagebreak
\widetext    
\setcounter{equation}{0}
\makeatletter 
\renewcommand{\theequation}{S\@arabic\c@equation}
\makeatother

\setcounter{figure}{0}
\makeatletter 
\renewcommand{\thefigure}{S\@arabic\c@figure}
\renewcommand{\bibnumfmt}[1]{[S#1]}
\renewcommand{\citenumfont}[1]{S#1}
\makeatother

\onecolumngrid

\begin{center}
{\bf \large Supplemental Material to ``Twins Percolation for Qubit Losses in Topological  Color Codes''}\\
\vspace{0.5cm}
{D. Vodola,$^{1}$ D. Amaro,$^{1}$ M.A. Martin-Delgado,$^{2}$ M. M{\"u}ller,$^{1}$  }\\
{\small
{$^1$}{\it Department  of  Physics,  Swansea  University,  Singleton  Park,  Swansea  SA2  8PP,  United  Kingdom.} \\
{$^2$}{\it Departamento de F\'isica Te\'orica I, Universidad Complutense, 28040 Madrid, Spain.} }
\end{center}

{\small

In this supplemental material we present important technical details as well as additional numerical results that are not shown in the main text, which underline the functioning and effectiveness of the qubit loss recovery algorithm for a variety of 2D color code lattices. 
Specifically, we start in Sec.~\ref{sec:ColorCodes} by providing background information on the 2D lattices and geometries we study numerically, and present in Sec.~\ref{sec:FindingLogicalOp} schematic examples of the different methods used to identify equivalent logical operators. In Sec.~\ref{sec:KitaevQubitLoss} we sketch the basic idea of protocol that has been proposed in Ref.~\cite{suppStace2009} for the correction of qubit losses in Kitaev's toric code, in order to highlight in more detail the fundamental differences with the protocol we introduce in the main text for the color code. In Sec.~\ref{Sec:RemovalExcitations} we provide the details of why excitations of the merged and reduced stabilizers as a result of the lattice reconstruction algorithm are correlated and can be removed deterministically and without the use of a decoding algorithm. Finally, the algorithm used to compute the qubit loss thresholds is described in more detail in Sec.~\ref{Sec:AlgorithmAndNumericalResults} and the numerical results obtained for 2D color codes defined on hexagonal (6.6.6) lattices, as well as for 6.6.6 and 4.8.8. color codes on planar triangular lattices, are presented and discussed. 
}

\begin{figure}[b]
	\centering
	\includegraphics[width=1\columnwidth]{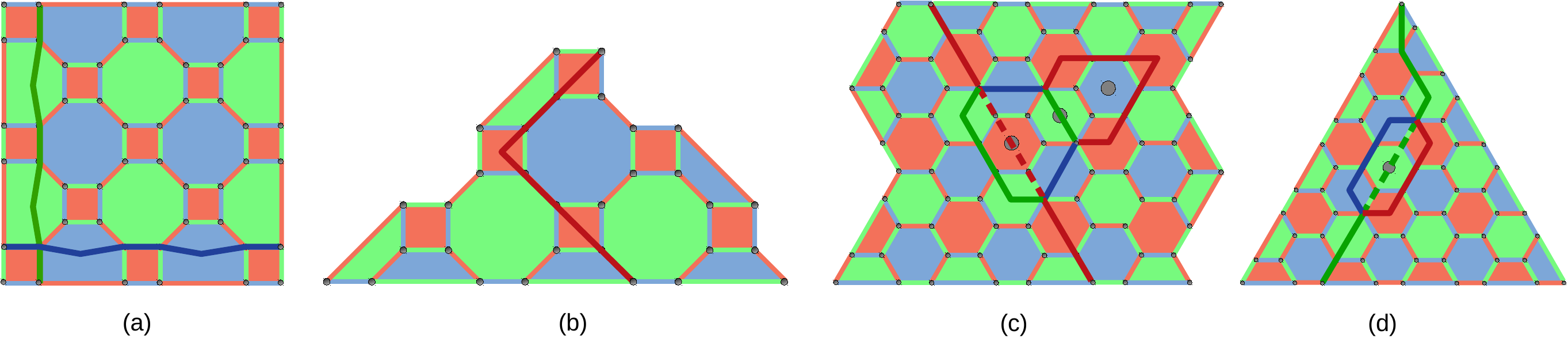}
	\caption{Color code lattices studied for the qubit loss problem. (a) Square 4.8.8. lattice of logical distance $ d=6 $ where the logical generators can be defined along a blue  and a green path traversing the lattice from the left to the right, and from the upper to the lower boundaries, correspondingly. (b) Triangular 4.8.8. lattice of distance $ d=7 $, with one red path. (c) Square 6.6.6 lattice of distance $ d=8 $ where the red path has a second level branching due to the action of the plaquettes marked with a grey circle. (d) Triangular 6.6.6. lattice of distance $ d=9 $ where the green path has a first level branching obtained by combining it with the green plaquette marked with a grey circle.}
	\label{lattices}
\end{figure}

\section{Color code and lattices}\label{sec:ColorCodes}
In this section we provide background information about color codes and about the lattices whose tolerance to qubit loss has been benchmarked: color codes defined on 4.8.8 and 6.6.6 lattices in their triangular and square versions [Fig.~\ref{lattices}].

As described in the main text, color codes \cite{suppBombin2006} are defined on 2D lattices where qubits are located on the vertices. The underlying lattices are necessarily three-colorable and trivalent, so the plaquettes and their edges can be colored with colors $ c=R,G,B $ in such a way that adjacent plaquettes have different colors $ c $, and the color of each edge is complementary to the two colors of the two plaquettes that share the edge. Instances of regular, translationally invariant lattices are the 4.8.8 lattice [Figs.~\ref{lattices}(a,b)], where plaquettes in the bulk are squares and octagons, and the 6.6.6 lattice [Figs.~\ref{lattices}(c,d)], i.e.~a honeycomb lattice. Whereas these are translationally invariant lattices, the qubit loss correction algorithm and associated protocols introduced in the main text also apply to more general lattices called colexes \cite{suppBombin2007}, which may not have any discrete symmetry.

The code space $ \mathcal{C} $ where logical qubits are encoded is determined by a set of stabilizer operators defined on the lattice. For each plaquette $ P $ there are two stabilizers: $ S_{P}^{\sigma}=\prod_{j\in P}\sigma_{j} $, where $ \sigma $ is a Pauli $ X $ or $ Z $ operator acting on all the qubits belonging to a given plaquette. All stabilizer operators mutually commute given that every plaquette contains an even number of qubits and different plaquettes share two or zero qubits. Thus, the stabilizer group $ \mathcal{S} $ is generated by the set of independent stabilizer generators and the code space $ \mathcal{C} $ is the common $ +1 $ eigenspace of all stabilizers.

The topology of the surface where the lattice is embedded uniquely determines the number $ k $ of logical qubits in $ \mathcal{C} $, independently of the system size and the bulk structure of the lattice hosting the code, via $ k=4-2\chi $ where $ \chi $ is the Euler characteristic of the surface. For instance, color codes defined on square lattices like the ones shown in Figs.~\ref{lattices}(a,c) encode two logical qubits, while the color codes on triangular lattices as the ones displayed in Figs.~\ref{lattices}(b,d) encode one logical qubit.

\begin{figure*}[t]
	\centering
	\includegraphics[width=1\columnwidth]{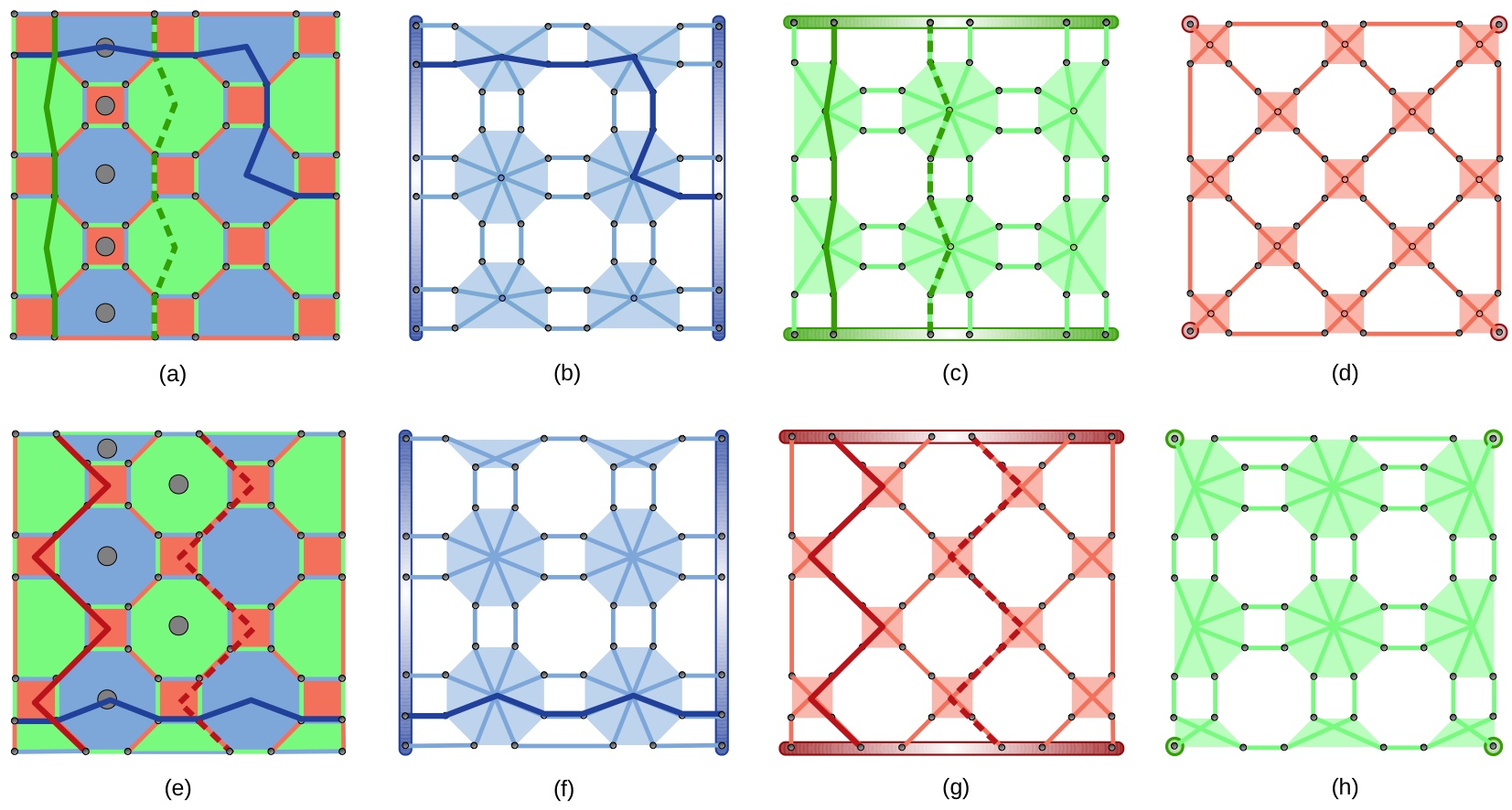}
	\caption{Two instances of a 4.8.8 square color code with distance $d=6$ and their shrunk lattices. (a) Square 4.8.8. color code lattice where blue and green logical operators are defined. The dashed green path has been deformed into the solid green line by the action of the plaquettes marked in grey. (b) Blue shrunk lattice with borders at the left and right sides of the lattice. The blue path goes from one border to the other. (c) Green shrunk lattice where borders are at the top and bottom sides of the lattice and the two greeen paths go from one to the other. (d) Red shrunk lattice. 
		(e-h) A different instance of a square 4.8.8. color code, where the logical operators belong to the blue and red shrunk lattices. We notice that in both instances in the blue and green shrunk lattices of a 4.8.8 color code, each plaquette is connected to another (or to a border) by two links, while for the red shrunk lattice each plaquette is connected to another (or to a border) by only one link. The different structure of the red lattice as compared to the blue and green shrunk lattices is related to a lower threshold for the existence of a logical operator given by a percolating string [method (I) in the main text)].}
	\label{shrunk}
\end{figure*}

In order to define the logical operators (generators) that generate the algebra of the encoded qubits it is convenient to introduce the concept of a shrunk lattice. A color code has three shrunk lattices, one for each color. For instance, the green shrunk lattice is obtained by placing a node at the centre of every green plaquette and connecting them through links [Figs.~\ref{shrunk}(c,h)]. Note that every link in the shrunk lattice corresponds to an edge of the original lattice of the same color. In planar codes, every shrunk lattice has at least two borders. The borders of a shrunk lattice consist of the qubits connected only through one link. For instance, the blue borders (the borders of the blue shrunk lattice) in Figs.~\ref{shrunk}(b,f) are the left and the right sides of the lattice. Borders can also consist of one qubit only. This is the case of the top right qubit in the triangular code of Fig.~\ref{lattices}(b) and of the four corners of the square lattice in Fig.~\ref{shrunk}(d). 

Having the definition of border in mind, we call any set of qubits that goes from one border to the other in the shrunk lattice of the color $ c $ [Fig.~\ref{shrunk}] a path $ Q_{c} $. All paths have in common that (1) they share an even number of qubits with every plaquette, that (2) they share zero or an even number of qubits with paths of the same color, and that (3) they share zero or an odd number of qubits with paths of a different color. 

For every encoded qubit $\mu$, we can now introduce a logical generator  $ T_{\mu}^{\sigma} $ defined on a path $ Q_{c} $ as $ T_{\mu}^{\sigma}=\prod_{i\in Q_{c}}\sigma_{i} $ for $ \sigma=X,Z $. Logical operators do not belong to $ \mathcal{S} $ but commute with the stabilizers because of (1) and they satisfy the required commutation and anticommutation relations among each other thanks to properties (2) and (3).

Importantly, two logical operators are equivalent if they differ by any element of the stabilizer $\mathcal{S}$, i.e., if there exists a subset $ \mathcal{V}\subseteq\mathcal{S} $ of stabilizer operators such that
\begin{equation}\label{eqn:ModifiedLogicalOperator}
	\tilde{T}_{\mu}^{\sigma}=T_{\mu}^{\sigma}\prod_{S\in\mathcal{V}}S.
\end{equation}
Equivalent logical operators satisfy the same (anti)commutation relations and have the same effect on the encoded data, but they have support on different sets of physical qubits $\tilde{Q}_{c}$. In particular, if a qubit $i$ is in $Q_{c}$ and a $\sigma$-stabilizer operator $ S_{P}^{\sigma} $ has support also on $i$, then $ \tilde{T}_{\mu}^{\sigma}={T}_{\mu}^{\sigma}S_{P}^{\sigma} $ will not have support on $i$ given that for Pauli operators $ \sigma^{2}$ is the identity.

In deforming a path  $Q_c$ of a given color $c$, two general cases are found by choosing different $\mathcal{V}$. If $\mathcal{V}$ is formed by plaquettes of color $c' \neq c$, the modified qubit set $ \tilde{Q}_{c}$ will still be a colored path inside the shrunk lattice of the same color $ c $ and we say that $ \tilde{Q}_{c}$  is a \textit{deformation} of $Q_c$. For instance, in Fig.~\ref{shrunk}(a), the green dashed path has been deformed by red and blue plaquettes marked with grey circles.

If instead $\mathcal{V}$ is made by plaquettes of the same color as $c$ we say that the path $ \tilde{Q}_{c} $ undergoes \textit{branching}. For example, a branched  green path splits into a red and a blue string, which belong to their respective shrunk lattices. We call that a first level of branching [see Fig.~\ref{lattices}(d)]. In the same way, red and blue strings can branch again, like in Fig.~\ref{lattices}(c), where the red string branches into a green and a blue string, and we call this a second level of branching. This freedom in choosing $\mathcal{V}$  has been used in the protocols described in the main text for finding a modified $\tilde{T}_{\mu}^{\sigma}$ that does not have support on lost and twin qubits.

\begin{figure}[b]
	\centering
	\includegraphics[width=0.65\textwidth]{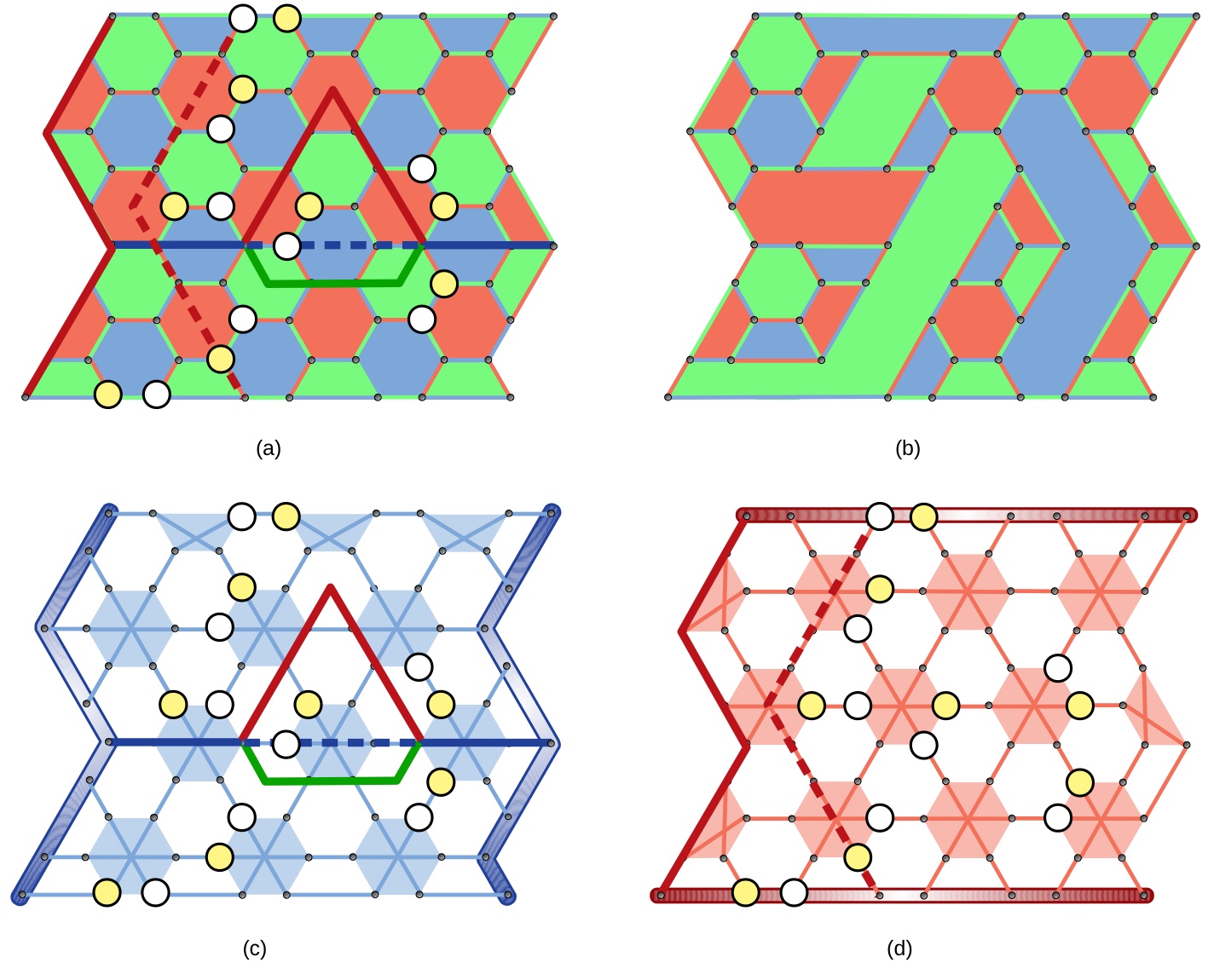}
	\caption{Square 6.6.6. lattice undergoing the loss of the qubits marked by white circles. (a) In this particular example, for each qubit that is lost, a twin qubit marked by a yellow circle is chosen and also removed from the lattice. The dashed red and blue paths contain missing qubits, so the red path can be deformed into the solid red line and the blue can undergo branching to avoid the missing qubits. (b) Reconstructed lattice obtained with the protocol described in the main text. (c) Blue shrunk lattice. One can check that there is no blue path that goes from one border to the other without touching a missing qubit, which means that the blue shrunk lattice does not percolate. However, a blue logical operator manages to reach the other border by branching into the red and green shrunk lattices, and thereby without visiting any of the missing qubits (losses and twin qubits). (d) Red shrunk lattice where the red dashed path contains two missing qubits, but it can be deformed into the red solid path, which does not contain any losses or twin qubits. Therefore, for the shown scenario, this red shrunk lattice percolates.}
	\label{losses}
\end{figure}

\section{Existence checks for equivalent logical operators}\label{sec:FindingLogicalOp}
In this section we provide details about the three ways (I), (II) and (III) described in the main text to find a modified logical operator $\tilde{T}_{\mu}^{\sigma}$ by Eq.~\eqref{eqn:ModifiedLogicalOperator} such that it does not have support on the set $ M $ of lost and twin qubits. We notice first that multiplying $ T_{\mu}^{\sigma} $ by stabilizers $ S_{P}^{\sigma'} $ of the other type $ \sigma'\neq\sigma $ is of no help for avoiding missing qubits. Therefore, it suffices to consider only the multiplication with stabilizers of the same type. In this way we can simply consider the path  $Q_c$ that defines the logical operator ${T}_{\mu}^{\sigma}$ and conclude that ${T}_{\mu}^{\sigma}$ is still defined if there exists a set of plaquettes $\set{P_j}$ such that the modified path $\tilde{Q}_c$ obtained by
\begin{equation}\label{eqn:FindEquivalentPath}
\tilde{Q}_c = Q_c \bigoplus_j P_j
\end{equation}
does not include any missing qubits. The symbol $\oplus$ denotes the symmetric difference, that is defined as $ A\oplus B \eqdef (A\cup B) \setminus (A\cap B) $ on sets $A$ and $B$. It is introduced because, for $\sigma$ Pauli operators, $\sigma^{2}$ is the identity and thus it ensures that $\tilde{Q}_c$ does not contain qubits that appear an even number of times in the collection $\set{Q_c, P_j}$. 

In Sec.~\ref{sec:ColorCodes}, we have seen that deformation and branching are the two strategies that paths have in order to avoid missing qubits. For instance, in Fig.~\ref{losses}(d), the red path is deformed to avoid the two losses that fall on it, while in Fig.~\ref{losses}(c), the blue path must branch to reach the other border. In order to study the robustness of color codes against losses, these two strategies were studied in three different methods described in the main text: (I) simple deformation, i.e., when paths of a color $c$ can be deformed  only inside the shrunk lattice of the same color, which is equivalent to a standard percolation check; (II) deformation and simple branching, i.e., situations in which paths of a color $c$ can be modified by only the first and second level of branching; (III) searching exhaustively by use of Eq.~\eqref{eqn:FindEquivalentPath} also among all higher-level deformations and branching process, and therefore among all equivalent paths generated by the action of the whole stabilizer group $\mathcal{S}$. 

Even though the number of equivalent paths grows exponentially with the number $ n $ of plaquettes, the search in (III) can be efficiently performed by mapping the plaquette sets $\{ P_j\} $ and the paths $ Q_{c} $ into binary matrices and checking the solvability of a system of linear equations that we are going to introduce below. To this end, recall that $N$ is the number of physical qubits, $n$ is the number of independent plaquettes and
\begin{itemize}
\item let $\mathbb{Q}_{c} =  (q_1,\dots,q_N)^T$ be a binary column vector where $q_i \in \{0,1\}$ is chosen such that $q_i = 1$ if the physical qubit $i$ appears in  the path $Q_{c}$, otherwise $q_i = 0$;
\item let $\mathbb{P}_{j} = (p_{j\,1},\dots,p_{j\,N})^T $ be a binary column vector for $j=1,\dots,n$ with $p_{j\,i} \in \{0,1\}$ and $p_{j\,i} = 1$ if the physical qubit $i$ appears in the plaquette $P_{j}$, otherwise $p_{j\,i} = 0$;
\item let $\mathbb{A} $ be a $ N\times n $ binary matrix whose $j$-th column is equal to the column vector~$ \mathbb{P}_{j} $;
\item let $\mathbb{M} = (m_1,\dots,m_N)^T$ be a binary column vector where $ m_{i}=1 $ if the $ i $-th qubit is a lost or a twin qubit, otherwise $ m_{i} = 0$. 
\end{itemize}

The symmetric difference between $ Q_{c} $ and $ P_{j} $ is then mapped, for the binary vectors, to $\mathbb{Q}_{c}\oplus\mathbb{P}_{j} $ where $\oplus$ stands for sum modulo~2 and all the operations among binary matrices and vectors will be also performed modulo 2. Solving Eq.~\eqref{eqn:FindEquivalentPath} is then equivalent to finding a row binary vector $x$ such that the binary column $\tilde{\mathbb{Q}}_{c}=\mathbb{A}x\oplus\mathbb{Q}_{c}$ that represents the modified path $\tilde{{Q}}_{c}$ will satisfy $ \mathbb{M}\circ \tilde{\mathbb{Q}}_{c} = 0$, where  the symbol $\circ$ denotes element-wise multiplication. This  can be translated into the linear system 
\begin{equation}\label{eqn:AlgebraicMethod}
(\mathbb{M}\circ \mathbb{A})x=\mathbb{M}\circ \mathbb{Q}_{c}
\end{equation}
where the existence of solutions for $x$ can be found efficiently by by standard Gaussian elimination algorithms, for which the runtime scales as $\sim N^3$ or less.

\begin{figure}[b]
	\centering
	\includegraphics[width=0.8\textwidth]{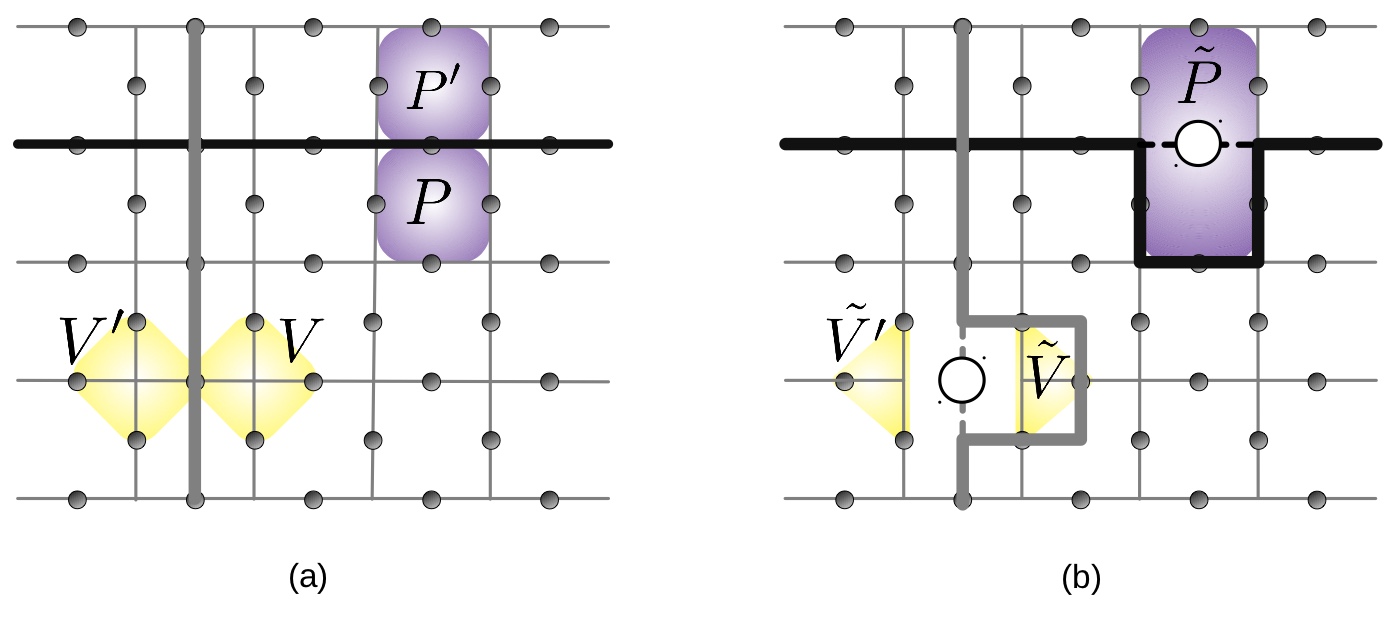}
	\caption{Lattice structure of Kitaev's toric or surface code. (a) Physical qubits (gray circles) reside on the links of a square lattice and the stabilizers are defined on the plaquettes and vertices. Plaquette stabilizers are product of Pauli $ Z $ operators acting on the qubits in the plaquettes like $ P $ and $ P' $ marked in purple. Vertex stabilizers act with Pauli $ X $ operators on sets  of neighbouring qubits (like $ V $ and $ V' $ marked in yellow). The $ Z $-type logical operator is defined on the black path going from the left to the right border and the $ X $-type logical operator is defined on the grey path going from the top to the bottom border. (b) According to the algorithm put forward by Stace \textit{et al.} \cite{suppStace2009}, the lattice hosting the toric/surface code can be reconstructed by merging the plaquettes $ P,P' $ sharing a lost qubit into a super-plaquette $ \tilde{P} $ like the one marked in purple, and modifying the vertex sets $ V,V' $ into $ \tilde{V},\tilde{V'} $ respectively to avoid the lost qubit. The $ X $ and $ Z $ logical operators are deformed by the action of the plaquette $ P $ and the vertex $ V $ respectively to avoid the positions (edges) corresponding to losses.}
	\label{fig:KitaevCode}
\end{figure}

\section{Qubit loss in the Kitaev surface code}\label{sec:KitaevQubitLoss}

In this section we briefly summarize main elements of the protocol pioneered in Refs.~\cite{suppStace2009,suppStace2010} to deal with qubit losses in the Kitaev code~\cite{suppKitaev2003}. We recall that in Kitaev's toric or surface code [Fig.~\ref{fig:KitaevCode}(a)] the physical qubits reside on the edges of a lattice, the stabilizer group is generated by four-qubit $Z$-type plaquette (purple) and $X$-type vertex (yellow) operators and the logical operators correspond to product of $Z$ (black) and $X$ (gray) operator along non-trivial strings extending across the entire lattice. 

Consider the lattice in Fig.~\ref{fig:KitaevCode}(b), which is damaged by two losses marked as white circles. For each lost qubit $q$, which is shared by two plaquettes $ P $ and $ P' $, a new so-called super-plaquette $ \tilde{P}=P\oplus P' $ that does not contain the lost qubit can be introduced, instead of $ P $ and $ P' $. The symbol $ \oplus $ indicates the symmetric difference between sets $ A\oplus B \eqdef (A\cup B) \setminus (A\cap B)$. Additionally, the two vertex operators [$ V $ and $ V' $ shown in yellow in Fig.~\ref{fig:KitaevCode}(a)], originally containing the lost qubit $ q $, shrink and are converted into three-qubit $X$-type stabilizers defined on the qubit sets $ \tilde{V}=V\setminus q $ and $ \tilde{V'}=V'\setminus q $ that do not contain $ q $. Overall, this defines a complete set of new stabilizers on the damaged lattice.

If a logical operator, like the one depicted in~Fig.~\ref{fig:KitaevCode}(a) with a solid line, has support on a loss, it can be deformed by the action of one or more plaquettes [in Fig.~\ref{fig:KitaevCode}(b) it is deformed by $ P $] such that it percolates through the entire lattice. In the case of the grey logical operator, it is deformed by the action of the vertex set $ V $. This procedure will fail if finding a percolating path is not possible anymore. Note that logical operators never undergo branching in the surface/toric code. Therefore, the boundary between recoverable and non-recoverable losses is given in this case by the bond percolation threshold on a 2D square lattice, that is known to be $p_c = 1/2$~\cite{suppStauffer1985}.

\section{Removal of excitations and modified stabilizer group}\label{Sec:RemovalExcitations}

\begin{figure}[b]\centering
\includegraphics[scale=0.4]{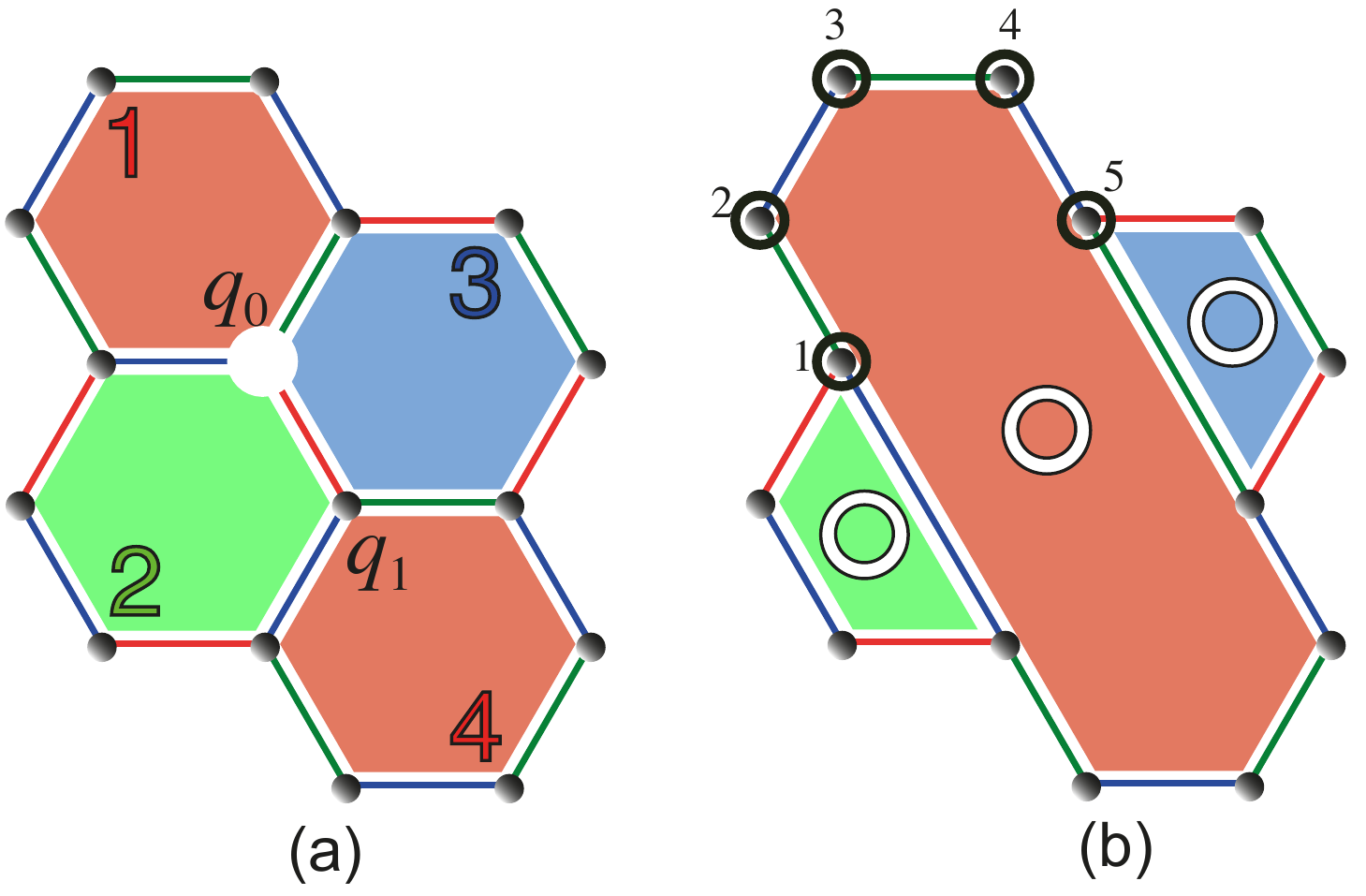}
\caption{(a) Subset $\Sigma$ of the stabilizer group $\mathcal{S}$ showing four $Z$ and $X$-stabilizers. The lost qubit $q_0$ is depicted as a white circle and the twin qubit is $q_1$. (b) After the lattice is reconstructed the plaquettes are in an undefined  state and excitations (represented as hollow circles) can be present and must be removed. One possible correction for bringing the system back into the code space is a product of Pauli operators acting on the qubits shown as black circles: this string of Pauli operators shares an odd number of qubits with each of the three new plaquettes, therefore it anticommutes with the associated $X$/$Z$ stabilizer generators of the complementary type $Z$/$X$, respectively.}
\label{fig:ExcitationRemoval}
\end{figure}

After our algorithm for dealing with losses has been applied and the existence of logical operators has been checked so that the logical information is not corrupted [steps (i, ii, iii) of the algorithm as outlined in the main text], the code will be in an undetermined state as it might be affected by excitations (i.e. $-1$ eigenstates of the newly defined merged or shrunk $X$ and $Z$-type stabilizer generators) that need to be removed. In this section, we will show how the excitations affect the stabilizers group~$\mathcal{S}$ and what will be new stabilizers~$\mathcal{S}'$ after removing them. One example of a reconstructed lattice affected by multiple losses is depicted in Fig.~\ref{losses}(b). We will describe the case of one single qubit loss, as it is straightforward to generalize it to more than one lost qubit. 

Consider the subset $\Sigma = \set{B^{X}_P, B^{Z}_P}_{P=1\dots 4} \subset  \mathcal{S}$ for the section of the code reported in Fig.~\ref{fig:ExcitationRemoval}(a) and, say, the couple lost-twin qubits is represented by the pair $(q_0,q_1)$. In order to end up in a valid code space $\mathcal{C}$ defined by independent and mutually commuting stabilizers, we have to build a new set $\Sigma'$  that has no support on $q_0$ and $q_1$. Let us define dimer operators $D^\sigma = \sigma_{q_0} \sigma_{q_1}$ with $\sigma=X,Z$. Since $D^X$ anticommutes with  $B^{Z}_1$ and  $B^{Z}_4$, its eigenvalue $\epsilon_X = \pm 1$ is undetermined and for updating the generator set we replace $B^{Z}_1$ with the product  $B^{Z}_1 B^{Z}_4$ and  $B^{Z}_4$ with  ${\epsilon_{X}  D^X}$. In the same way, the operator $D^Z= Z_{q_0} Z_{q_1}$ anticommutes with  $B^{X}_1$ and  $B^{X}_4$, so we replace $B^{X}_1$  with the product  $B^{X}_1 B^{X}_4$ and $B^{X}_4$ with ${\epsilon_{Z}  D^Z}$. We can now use the dimer operators $D^X, D^Z$ to decouple the generators from the qubits $q_0$ and $q_1$. To do so, we multiply all the $B^{\sigma}_P$ that contain $q_0$ and $q_1$ by  $\epsilon_\sigma D^\sigma$. The final set of generators is
\begin{equation}
\Sigma = \set{\epsilon_\sigma D^\sigma, \epsilon_\sigma D^\sigma B^\sigma_1 B^\sigma_4, \epsilon_\sigma D^\sigma  B^\sigma_2, \epsilon_\sigma D^\sigma B^\sigma_3, \sigma= X,Z}
\end{equation}
and corresponds to the lattice reconstructed by the steps  (i, ii) as well as the dimer formed by the lost and the twin qubit. This set of generators $\Sigma$ will be then split into two disjoint subsets, one containing the two dimer operators $\set{\epsilon_\sigma D^\sigma}$, the other $\Sigma'$ having support on the remaining qubits of the reconstructed lattice [Fig.~\ref{fig:ExcitationRemoval}(b)]. The set  $\Sigma'$ contains generators that are in an undetermined state. In order to fix the state of the code, as all the three redefined $\sigma$-type generators show correlated excitations, one has to measure only one generator for each $\sigma$ and thus detect whether an excitation $\epsilon_\sigma = -1$ is present. If it is the case, we then need to remove the excitations by bringing the state back to the code space. For example, if $\epsilon_Z = -1$, an $X$-type excitation will affect the new-defined plaquettes and one possible correction will be the product of $X$-Pauli operators on the three qubits as Fig.~\ref{fig:ExcitationRemoval}(b) shows. For more than a single loss, this argument can be easily iterated, with several binary unknowns $\epsilon$ being introduced, one for each of removed dimer. 

\section{Algorithm for computing the critical loss rate and numerical results}\label{Sec:AlgorithmAndNumericalResults}

In this section we explain in more detail the algorithm used to obtain the loss thresholds at finite code distance $d$ and in the thermodynamic limit $d\to\infty$. We also present the results for the triangular and the 6.6.6 lattices that are not discussed in the main text. For each of the lattices in Fig.~\ref{lattices}, we fix a logical distance $ d $ and find the critical loss rate $ p_{c}(d) $ at which a logical operator can no longer be found according to each of the three methods explained in the main text and with more detail in Sec.~\ref{sec:FindingLogicalOp} of this Supplemental Material. 

\begin{figure}[b]
	\centering
	\includegraphics[width=0.6\textwidth]{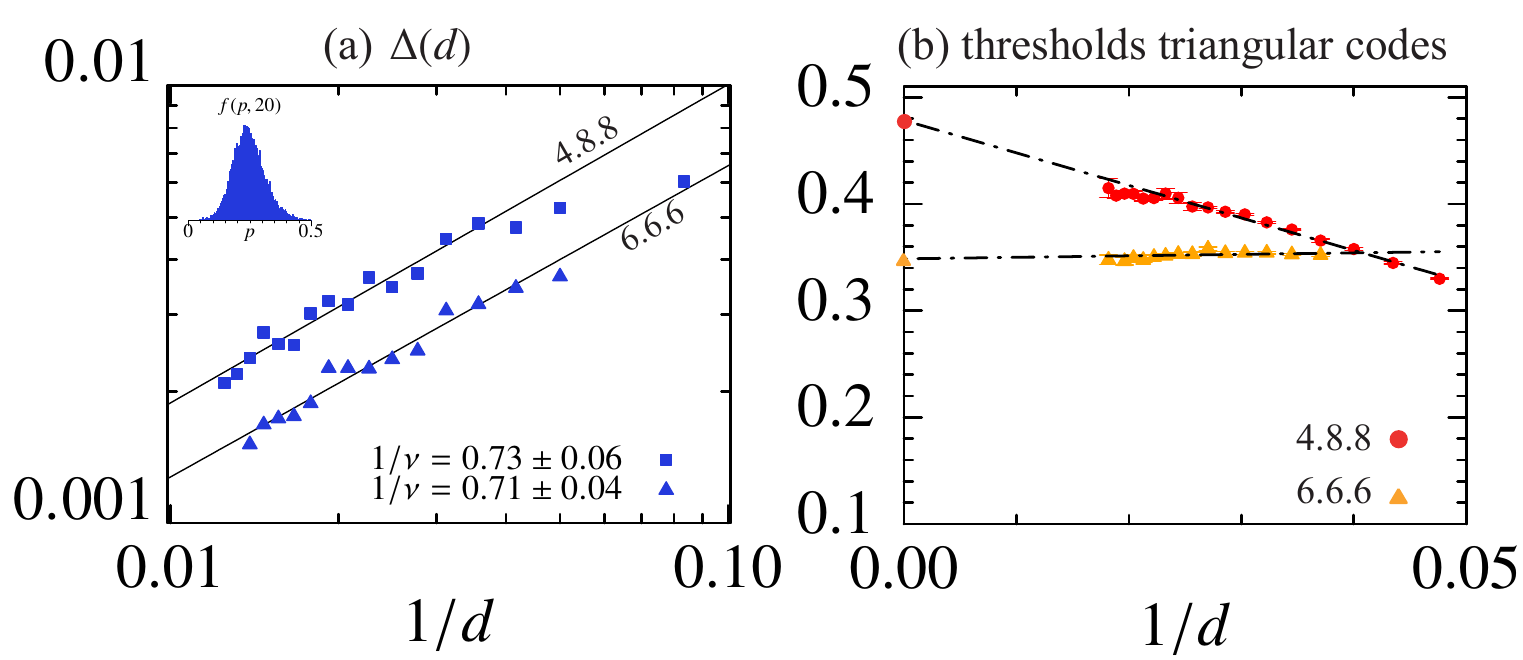}
	\caption{(a) Standard deviation $\Delta(d)$ of the distribution $f(p,d)$ as a function of $1/d$ for the blue shrunk lattices of the 4.8.8 and the 6.6.6 lattices. An example of $f(p,d)$ is represented in the inset for $d=20$. For both the shrunk lattices, $\Delta(d)\propto 1/d^{1/\nu}$ with $1/\nu\approx 3/4 $. The points are fitted by the solid lines scaling as $ 1/d^{3/4}$. (b) Critical threshold scaling in the limit $d\to\infty$ for the triangular 4.8.8 and 6.6.6 lattices computed by solving Eq.~(1) of the main text [equivalent to Eq.~\eqref{eqn:AlgebraicMethod}]. The values in the thermodynamic limit are represent by point on the vertical axis and coincide with the ones computed for the corresponding square lattice [See Fig.~3(a) of the main text and Fig.~\ref{fig:numerics_666}].}\label{fig:ScalingExp}
	
	\vspace{1cm}
	
	\includegraphics[width=0.85\textwidth]{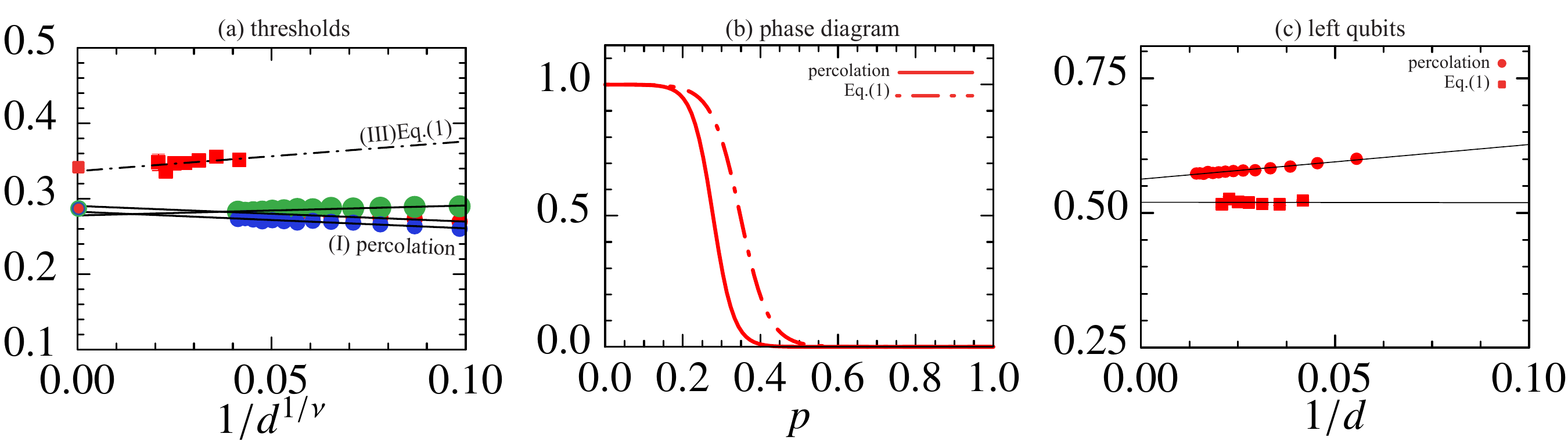}
	\caption{Numerical results for the critical threshold $p_\infty$ for the 6.6.6 square lattice. (a) Critical loss rate $ p_{c} $ as a function of the lattice distance $ d $. Circles represent the percolating critical loss rate for the three shrunk lattices (points are fitted by the solid lines scaling as $ 1/d^{3/4}$) while squares represent the fundamental thresholds from Eq.~(1) of the main text for the red shrunk lattice (points are fitted by the solid line scaling as $ 1/d$). (b) Phase diagram as a function of the loss rate $ p $ for a lattice with $d = 48$ representing the probability of finding a percolating path (solid line) and a solution of Eq.~(1) of the main text (dot-dashed line). (c) Fraction of left qubits as a function of $ 1/d $. Again, squares represent results obtained from the algebraic method of Eq.~(1)  and circles are obtained with the percolating method.}\label{fig:numerics_666}
\end{figure}

We start by drawing a set of losses randomly with a rate $ p_{1}=1/2 $. Then, we reconstruct the lattice by following the protocol explained in the main text and after that we check if a logical operator still exists. If it is the case, the process is repeated, but using a bigger loss rate $ p_{2}=p_{1}+1/4 $, otherwise, the loss rate is reduced by the same factor $ p_{2}=p_{1}-1/4 $. For each round, the loss rate is increased (reduced) by half of the previous factor if we can (cannot) find an equivalent logical operator. The algorithm scales only logarithmically with the number of qubits in the lattice and halts when the number of lost qubits in two successive rounds does not change anymore. This algorithms gives a distribution $f(p,d)$ of loss rates at which a logical operator is found for the first time for a given fixed code distance. An example $f(p,d)$ is plotted in the inset of Fig.~\ref{fig:ScalingExp}. The distribution $f(p,d)$  can also be interpreted as the probability that the logical operator ceases to be defined if the rate of losses is increased from $p$ to $p+\mathrm{d}p$. Finally, we take the mean value of $f(p,d)$ as the critical loss rate $ p_{c}(d) $ corresponding to the fixed distance $ d $.

Regarding the methods (I) and (II) described in the main text and in Sec.~\ref{sec:FindingLogicalOp}, in order to extract the value  $p_\infty$ of the threshold for $d\to\infty$, we use a finite-size scaling analysis borrowed from percolation theory~\cite{suppStauffer1985}. For $p$ near the critical value $p_\infty$, the correlation length $\xi$, characterizing the typical size of a path connecting two random plaquettes, is known to diverge as $\xi \propto |p - p_\infty|^{-\nu} $ for $d\to\infty$ with an exponent~$\nu$ that percolation theory predicts to depend only on the dimensionality of the system. All the quantities showing a scaling law near $p_\infty$ are controlled by the ratio $d/\xi$, i.e.~by a scaling variable $z = (p - p_\infty) d^{1/\nu}$. In particular, let $F(p,d)$ be the probability that a distance-$d$ code affected by a rate $p$ of losses shows a percolating path. Clearly, the distribution $f(p,d)$ we computed numerically satisfies $f(p,d) = \mathrm{d} F / \mathrm{d} p$.

For $d\to\infty$, $F$ will approach a step function with a discontinuity at $p_\infty$ that separates a region ($p<p_\infty$) where a logical operator is always found from another ($p>p_\infty$) where it does not exist, while $f$ will become a Dirac delta peaked around $p_\infty$. For sufficiently large, but finite $d$ and for $p$ close enough to $p_\infty$,  $F(p,d)$ is expected to  be a function only of $z$, i.e.$F(p,d) = F(z)$ and $f(p,d) = d^{1/\nu}\mathrm{d}F(z) /\mathrm{d} z$. From the last equation, one can see that $p_c(d)$, i.e.~the mean value of $f(p,d)$, differs from $p_\infty$ according to 
\begin{equation}\label{eqn:FSS_probability}
p_c(d) - p_\infty \propto \frac{1}{d^ {1/\nu}},
\end{equation} 
while the standard deviation of $f(p,d)$ will scale as $\Delta(d) \propto d^{-1/\nu}$. We use these last two equations to compute $\nu$ and $p_\infty$ and thereby to determine the thresholds for qubit losses for the color codes when using a deformed or a branched logical operator.

Figure~\ref{fig:ScalingExp}(a) shows $\Delta(d)$  as a function of $1/d$ in log-log scale for the blue shrunk lattice of the square 4.8.8 and the 6.6.6  codes. Similar plots are found for the other colored shrunk lattice. The slope of the curves gives the exponent $1/\nu$. Interestingly, one can see that $1/\nu$ for this generalized percolation problem takes the value of 3/4 expected for the standard percolation~\cite{suppStauffer1985}.

When instead we determine the threshold by solving Eq.~(1) of the main text [equivalent to Eq.~\eqref{eqn:AlgebraicMethod}], we also use Eq.~\eqref{eqn:FSS_probability}, but with an exponent $\nu=1$ as no natural scaling law and thus critical exponent are expected.

\subsection{Numerical results for the triangular codes and the 6.6.6 lattices}
For the 4.8.8 and the 6.6.6 triangular codes, Fig.~\ref{fig:ScalingExp}(b) shows the threshold rates obtained by solving Eq.~(1) of the main text [equivalent to Eq.~\eqref{eqn:AlgebraicMethod}]. The dot-dashed lines represent fits scaling as $p_\infty + b/d$. For both the lattices, $p_\infty$ can be read on the vertical axis.

Data for the 6.6.6 square lattice are reported in Fig.~\ref{fig:numerics_666}(a) that shows $p_c(d)$ as a function of $1/d^{1/\nu}$ computed via the methods (I) and (III) described in the main text and in Sec.~\ref{eqn:AlgebraicMethod}. Linear fits whose intercept yield $p_\infty$ are also drawn. As for the 4.8.8 lattice the method (III), the one based on checking whether the system in Eq.~(1) of the main text admits solutions, yields a higher threshold. For a lattice with $d = 48$, Fig.~\ref{fig:numerics_666}(b) shows the probability of finding a red logical operator by using method (I), i.e., simple percolation and method (III). It then represents the phase diagram describing the boundary separating a region where the qubit associated to the logical red operator can be successfully recovered from a region where it is completely destroyed. For completeness, Fig.~\ref{fig:numerics_666}(c) shows the fraction of qubits left in the lattice as a function of $1/d$. Notably, this number approaches the limiting value of 50\% when considering the solution of system of Eq.~(1), demonstrating the high robustness of color codes against qubit loss.

\end{document}